\documentclass[aps,prd,twocolumn,superscriptaddress,groupedaddress,nofootinbib]{revtex4}

\usepackage{amsmath,amssymb,bm,color,dcolumn,mathrsfs,tabularx,graphicx}
\usepackage{hyperref}
\hypersetup{colorlinks=true,linkcolor=blue,citecolor=blue,urlcolor=blue}
\usepackage{Macro} 

	\def\bicep{{\sc BICEP}}
	\def\bicepone{{\sc BICEP1}}
	\def\biceptwo{{\sc BICEP2}}
	\def\keckarray{{\it Keck Array}}

	\def\BK{\bicep/\keckarray}
	
	\def\bk{{\rm BK}14}

	\def\a{\alpha}

	\def\ACB{A_{\rm CB}}
	\def\hACB{\widehat{A}_{\rm CB}}

	\def\rE{E'}
	\def\rB{B'}

	\def\tCEE{\widetilde{C}^{\rm EE}}

	\def\l{\ell}

	\def\bl{\bm{\ell}}
	
	\def\bL{\textit{\textbf{L}}}
	
	\def\hatn{\hat{\textit{\textbf{n}}}}

	\def\eq#1{Eq.~\eqref{#1}}

\newcommand{\compile}{}

\begin{document}

\ifdefined\compile 

\title{\biceptwo\ / \keckarray\ IX: New Bounds on Anisotropies of CMB Polarization Rotation and Implications for Axionlike Particles and Primordial Magnetic Fields}


\author{\keckarray\ and \biceptwo\ Collaborations: 
P.~A.~R.~Ade}
\affiliation{School of Physics and Astronomy, Cardiff University, Cardiff, CF24 3AA, United Kingdom}
\author{Z.~Ahmed}
\affiliation{Kavli Institute for Particle Astrophysics and Cosmology, SLAC National Accelerator Laboratory, 2575 Sand Hill Rd, Menlo Park, California 94025, USA}
\affiliation{Department of Physics, Stanford University, Stanford, California 94305, USA}
\author{R.~W.~Aikin}
\affiliation{Department of Physics, California Institute of Technology, Pasadena, California 91125, USA}
\author{K.~D.~Alexander}
\affiliation{Harvard-Smithsonian Center for Astrophysics, 60 Garden Street MS 42, Cambridge, Massachusetts 02138, USA}
\author{D.~Barkats}
\affiliation{Harvard-Smithsonian Center for Astrophysics, 60 Garden Street MS 42, Cambridge, Massachusetts 02138, USA}
\author{S.~J.~Benton}
\affiliation{Department of Physics, University of Toronto, Toronto, Ontario, M5S 1A7, Canada}
\author{C.~A.~Bischoff}
\affiliation{Department of Physics, University of Cincinnati, Cincinnati, Ohio, 45221, USA}
\author{J.~J.~Bock}
\affiliation{Department of Physics, California Institute of Technology, Pasadena, California 91125, USA}
\affiliation{Jet Propulsion Laboratory, Pasadena, California 91109, USA}
\author{R.~Bowens-Rubin}
\affiliation{Harvard-Smithsonian Center for Astrophysics, 60 Garden Street MS 42, Cambridge, Massachusetts 02138, USA}
\author{J.~A.~Brevik}
\affiliation{Department of Physics, California Institute of Technology, Pasadena, California 91125, USA}
\author{I.~Buder}
\affiliation{Harvard-Smithsonian Center for Astrophysics, 60 Garden Street MS 42, Cambridge, Massachusetts 02138, USA}
\author{E.~Bullock}
\affiliation{Minnesota Institute for Astrophysics, University of Minnesota, Minneapolis, Minnesota 55455, USA}
\author{V.~Buza}
\affiliation{Harvard-Smithsonian Center for Astrophysics, 60 Garden Street MS 42, Cambridge, Massachusetts 02138, USA}
\affiliation{Department of Physics, Harvard University, Cambridge, MA 02138, USA}
\author{J.~Connors}
\affiliation{Harvard-Smithsonian Center for Astrophysics, 60 Garden Street MS 42, Cambridge, Massachusetts 02138, USA}
\author{B.~P.~Crill}
\affiliation{Jet Propulsion Laboratory, Pasadena, California 91109, USA}
\author{L.~Duband}
\affiliation{Service des Basses Temp\'{e}ratures, Commissariat \`{a} l'Energie Atomique, 38054 Grenoble, France}
\author{C.~Dvorkin}
\affiliation{Department of Physics, Harvard University, Cambridge, MA 02138, USA}
\author{J.~P.~Filippini}
\affiliation{Department of Physics, California Institute of Technology, Pasadena, California 91125, USA}
\affiliation{Department of Physics, University of Illinois at Urbana-Champaign, Urbana, Illinois 61801, USA}
\author{S.~Fliescher}
\affiliation{School of Physics and Astronomy, University of Minnesota, Minneapolis, Minnesota 55455, USA}
\author{T.~St.~Germaine}
\affiliation{Department of Physics, Harvard University, Cambridge, MA 02138, USA}
\author{T.~Ghosh}
\affiliation{Department of Physics, California Institute of Technology, Pasadena, California 91125, USA}
\author{J.~Grayson}
\affiliation{Department of Physics, Stanford University, Stanford, California 94305, USA}
\author{S.~Harrison}
\affiliation{Harvard-Smithsonian Center for Astrophysics, 60 Garden Street MS 42, Cambridge, Massachusetts 02138, USA}
\author{S.~R.~Hildebrandt}
\affiliation{Department of Physics, California Institute of Technology, Pasadena, California 91125, USA}
\affiliation{Jet Propulsion Laboratory, Pasadena, California 91109, USA}
\author{G.~C.~Hilton}
\affiliation{National Institute of Standards and Technology, Boulder, Colorado 80305, USA}
\author{H.~Hui}
\affiliation{Department of Physics, California Institute of Technology, Pasadena, California 91125, USA}
\author{K.~D.~Irwin}
\affiliation{Department of Physics, Stanford University, Stanford, California 94305, USA}
\affiliation{Kavli Institute for Particle Astrophysics and Cosmology, SLAC National Accelerator Laboratory, 2575 Sand Hill Rd, Menlo Park, California 94025, USA}
\affiliation{National Institute of Standards and Technology, Boulder, Colorado 80305, USA}
\author{J.~Kang}
\affiliation{Department of Physics, Stanford University, Stanford, California 94305, USA}
\affiliation{Kavli Institute for Particle Astrophysics and Cosmology, SLAC National Accelerator Laboratory, 2575 Sand Hill Rd, Menlo Park, California 94025, USA}
\author{K.~S.~Karkare}
\affiliation{Harvard-Smithsonian Center for Astrophysics, 60 Garden Street MS 42, Cambridge, Massachusetts 02138, USA}
\author{E.~Karpel}
\affiliation{Department of Physics, Stanford University, Stanford, California 94305, USA}
\author{J.~P.~Kaufman}
\affiliation{Department of Physics, University of California, San Diego, La Jolla, California 92093, USA}
\author{B.~G.~Keating}
\affiliation{Department of Physics, University of California, San Diego, La Jolla, California 92093, USA}
\author{S.~Kefeli}
\affiliation{Department of Physics, California Institute of Technology, Pasadena, California 91125, USA}
\author{S.~A.~Kernasovskiy}
\affiliation{Department of Physics, Stanford University, Stanford, California 94305, USA}
\author{J.~M.~Kovac}
\affiliation{Harvard-Smithsonian Center for Astrophysics, 60 Garden Street MS 42, Cambridge, Massachusetts 02138, USA}
\affiliation{Department of Physics, Harvard University, Cambridge, MA 02138, USA}
\author{C.~L.~Kuo}
\affiliation{Department of Physics, Stanford University, Stanford, California 94305, USA}
\affiliation{Kavli Institute for Particle Astrophysics and Cosmology, SLAC National Accelerator Laboratory, 2575 Sand Hill Rd, Menlo Park, California 94025, USA}
\author{N.~A.~~Larsen}
\affiliation{Department of Physics, Enrico Fermi Institute, University of Chicago, Chicago, IL 60637, USA}
\author{E.~M.~Leitch}
\affiliation{Kavli Institute for Cosmological Physics, University of Chicago, Chicago, IL 60637, USA}
\author{K.~G.~Megerian}
\affiliation{Jet Propulsion Laboratory, Pasadena, California 91109, USA}
\author{L.~Moncelsi}
\affiliation{Department of Physics, California Institute of Technology, Pasadena, California 91125, USA}
\author{T.~Namikawa$^\dagger$}
\affiliation{Department of Physics, Stanford University, Stanford, California 94305, USA}
\affiliation{Kavli Institute for Particle Astrophysics and Cosmology, SLAC National Accelerator Laboratory, 2575 Sand Hill Rd, Menlo Park, California 94025, USA}
\author{C.~B.~Netterfield}
\affiliation{Department of Physics, University of Toronto, Toronto, Ontario, M5S 1A7, Canada}
\affiliation{Canadian Institute for Advanced Research, Toronto, Ontario, M5G 1Z8, Canada}
\author{H.~T.~Nguyen}
\affiliation{Jet Propulsion Laboratory, Pasadena, California 91109, USA}
\author{R.~O'Brient}
\affiliation{Department of Physics, California Institute of Technology, Pasadena, California 91125, USA}
\affiliation{Jet Propulsion Laboratory, Pasadena, California 91109, USA}
\author{R.~W.~Ogburn~IV}
\affiliation{Department of Physics, Stanford University, Stanford, California 94305, USA}
\affiliation{Kavli Institute for Particle Astrophysics and Cosmology, SLAC National Accelerator Laboratory, 2575 Sand Hill Rd, Menlo Park, California 94025, USA}
\author{C.~Pryke}
\affiliation{School of Physics and Astronomy, University of Minnesota, Minneapolis, Minnesota 55455, USA}
\affiliation{Minnesota Institute for Astrophysics, University of Minnesota, Minneapolis, Minnesota 55455, USA}
\author{S.~Richter}
\affiliation{Harvard-Smithsonian Center for Astrophysics, 60 Garden Street MS 42, Cambridge, Massachusetts 02138, USA}
\author{A.~Schillaci}
\affiliation{Department of Physics, California Institute of Technology, Pasadena, California 91125, USA}
\author{R.~Schwarz}
\affiliation{School of Physics and Astronomy, University of Minnesota, Minneapolis, Minnesota 55455, USA}
\author{C.~D.~Sheehy}
\affiliation{Brookhaven National Laboratory, Upton, NY 11973, USA}
\author{Z.~K.~Staniszewski}
\affiliation{Department of Physics, California Institute of Technology, Pasadena, California 91125, USA}
\affiliation{Jet Propulsion Laboratory, Pasadena, California 91109, USA}
\author{B.~Steinbach}
\affiliation{Department of Physics, California Institute of Technology, Pasadena, California 91125, USA}
\author{R.~V.~Sudiwala}
\affiliation{School of Physics and Astronomy, Cardiff University, Cardiff, CF24 3AA, United Kingdom}
\author{G.~P.~Teply}
\affiliation{Department of Physics, University of California, San Diego, La Jolla, California 92093, USA}
\author{K.~L.~Thompson}
\affiliation{Department of Physics, Stanford University, Stanford, California 94305, USA}
\affiliation{Kavli Institute for Particle Astrophysics and Cosmology, SLAC National Accelerator Laboratory, 2575 Sand Hill Rd, Menlo Park, California 94025, USA}
\author{J.~E.~Tolan}
\affiliation{Department of Physics, Stanford University, Stanford, California 94305, USA}
\author{C.~Tucker}
\affiliation{School of Physics and Astronomy, Cardiff University, Cardiff, CF24 3AA, United Kingdom}
\author{A.~D.~Turner}
\affiliation{Jet Propulsion Laboratory, Pasadena, California 91109, USA}
\author{A.~G.~Vieregg}
\affiliation{Harvard-Smithsonian Center for Astrophysics, 60 Garden Street MS 42, Cambridge, Massachusetts 02138, USA}
\affiliation{Department of Physics, Enrico Fermi Institute, University of Chicago, Chicago, IL 60637, USA}
\affiliation{Kavli Institute for Cosmological Physics, University of Chicago, Chicago, IL 60637, USA}
\author{A.~C.~Weber}
\affiliation{Jet Propulsion Laboratory, Pasadena, California 91109, USA}
\author{D.~V.~Wiebe}
\affiliation{Department of Physics and Astronomy, University of British Columbia, Vancouver, British Columbia, V6T 1Z1, Canada}
\author{J.~Willmert}
\affiliation{School of Physics and Astronomy, University of Minnesota, Minneapolis, Minnesota 55455, USA}
\author{C.~L.~Wong}
\affiliation{Harvard-Smithsonian Center for Astrophysics, 60 Garden Street MS 42, Cambridge, Massachusetts 02138, USA}
\affiliation{Department of Physics, Harvard University, Cambridge, MA 02138, USA}
\author{W.~L.~K.~Wu}
\affiliation{Department of Physics, Stanford University, Stanford, California 94305, USA}
\affiliation{Department of Physics, University of California, Berkeley, CA 94720, USA}
\author{K.~W.~Yoon}
\affiliation{Department of Physics, Stanford University, Stanford, California 94305, USA}
\affiliation{Kavli Institute for Particle Astrophysics and Cosmology, SLAC National Accelerator Laboratory, 2575 Sand Hill Rd, Menlo Park, California 94025, USA}

\date{\today}


\begin{abstract}
We present the strongest constraints to date on anisotropies of cosmic microwave background (CMB) 
polarization rotation derived from $150$ GHz data taken by the \biceptwo\ \& \keckarray\ CMB 
experiments up to and including the 2014 observing season (\bk). 
The definition of the polarization angle in \bk\ maps has gone through self-calibration 
in which the overall angle is adjusted to minimize the observed $TB$
and $EB$ power spectra. After this procedure, the $QU$ maps lose sensitivity to a uniform 
polarization rotation but are still sensitive to anisotropies of polarization rotation. This analysis 
places constraints on the anisotropies of polarization rotation, which could be generated by CMB photons interacting 
with axionlike pseudoscalar fields or Faraday rotation induced by primordial magnetic fields. 
The sensitivity of \bk\ maps ($\sim 3\mu$K-arcmin) makes it possible
to reconstruct anisotropies of the polarization rotation angle and measure their angular power spectrum much
more precisely than previous attempts. Our data are found to be consistent with no polarization rotation
anisotropies, improving the upper bound on the amplitude of the rotation angle spectrum
by roughly an order of magnitude compared to the previous best constraints.
Our results lead to an order of magnitude better constraint on the coupling constant of the Chern-Simons electromagnetic term 
$g_{a\gamma}\leq 7.2\times 10^{-2}/H_I$ ($95\%$ confidence) than the constraint derived 
from the $B$-mode spectrum, where $H_I$ is the inflationary Hubble scale.
This constraint leads to a limit on the decay constant of $10^{-6}\alt f_a/M_{\rm pl}$
at mass range of $10^{-33}\leq m_a\leq 10^{-28}$ eV for $r=0.01$, 
assuming $g_{a\gamma}\sim\alpha/(2\pi f_a)$ with $\alpha$ denoting the fine structure constant. 
The upper bound on the amplitude of the primordial magnetic fields is $30$nG ($95\%$ confidence) 
from the polarization rotation anisotropies. 
\end{abstract}

\maketitle

\fi 

\renewcommand{\thefootnote}{\fnsymbol{footnote}}
\footnote[0]{$^\dagger$ Corresponding author: T.~Namikawa, 
\href{toshiyan@stanford.edu}{toshiyan@stanford.edu}}

\section{Introduction}

The \bicep/\keckarray\ (BK) program has been making deep observations of cosmic microwave background (CMB)
polarization at the South Pole. The $150$ GHz data taken through 2014 (BK14) have been used to 
constrain primordial gravitational waves (GWs) to $r<0.07$ ($95\%$ confidence, including Planck and WMAP) \cite{BKVI} 
and to detect gravitational lensing with high significance \cite{BKVIII}. 

In addition to GWs and lensing, CMB polarization can also be used to test various theories of 
physics beyond the Standard Model. Measurements of the polarization rotation angle are known to be a unique probe of 
new physics containing pseudoscalar fields
coupled with photons through the Chern-Simons term \cite{Ni:1977,Carroll:1998,Li:2008,Pospelov:2009,Finelli:2009,Caldwell:2011,Liu:2016dcg}:
\al{
	\mC{L} \supset \frac{g_{a\gamma}a}{4}F_{\mu\nu}\widetilde{F}^{\mu\nu} \,.
}
Here $a$ is a pseudoscalar field, $g_{a\gamma}$ is corresponding coupling constant, $F_{\mu\nu}$ is 
the electromagnetic field, and $\tilde{F}^{\mu\nu}$ is the dual of $F^{\mu\nu}$ 
(for review see e.g. Ref.~\cite{Marsh:2016} and references therein). 
The existence of the above pseudoscalar fields, also known as axionlike particles, 
is a generic prediction of string theory, 
and detection or any constraints on these fields can provide valuable implications for fundamental physics.
The presence of the above pseudoscalar fields leads to cosmic birefringence, in which 
the CMB polarization angle is rotated by 
\al{
	\a = \frac{\Delta a g_{a\gamma}}{2} \,, \label{Eq:alpha}
}
where $\Delta a$ is the change of the pseudoscalar fields along the photons' trajectory between 
the observer and recombination (e.g. Ref.~\cite{Pospelov:2009}). 
Fluctuations in $\Delta a$, as some models predict, lead to spatial variations in $\a$ 
(e.g. Refs.~\cite{Kamionkowski:2010,Caldwell:2011,Gluscevic:2012qv,Leon:2017}). 
If the pseudoscalar field is effectively massless during inflation, the power spectrum of 
the fluctuations of the pseudoscalar field has a scale-invariant spectrum. 
Ref.~\cite{Caldwell:2011} shows that the power spectrum of $\a$ induced by these fluctuations 
is given as 
\al{
	\sqrt{\frac{L(L+1)C_L^{\a\a}}{2\pi}} = \frac{H_Ig_{a\gamma}}{4\pi}  \,, \label{Eq:scaleinv}
}
in the large-scale limit ($L\alt 100$). Here $H_I$ is the inflationary Hubble parameter. 
Henceforth, we use $L$ for the multipoles of $\a$ and $\l$ for the CMB $E$ and $B$ modes. 

The measurements of the rotation angle can also be used to probe primordial magnetic fields (PMFs) 
through the Faraday rotation of CMB polarization \cite{Kosowsky:1996,Harari:1997}. 
In the large-scale limit ($L\alt 100$), nearly scale-invariant PMFs lead to \cite{Yadav:2012b,De:2013}
\al{
	\sqrt{\frac{L(L+1)C_L^{\a\a}}{2\pi}} = 1.9\times 10^{-4}\left(\frac{\nu}{150\,{\rm GHz}}\right)^{-2}
		\left(\frac{B_{1{\rm Mpc}}}{1\,{\rm nG}}\right) \,. \label{Eq:pmfs}
}
The rotation angle from PMFs depends on the observing frequency. 
Compared to the BK14 150 GHz data, the BK14 95 GHz data have larger noise and lower angular resolution, 
and the 150 GHz data from BK14 place the strongest constraints on PMFs. 
Thus we use the 150 GHz data in the following analysis. 
If we were to detect a rotation signal, then we would look for the same signal at $95$ GHz to test 
whether it has the correct wavelength dependence for Faraday rotation. 

The polarization rotation effect modifies the pattern of the CMB polarization map and leads to mixing of $E$ and $B$ modes. 
Since $E$ modes at last scattering are much brighter than the $B$ mode, this effect is mostly characterized by 
leakage from $E$ to $B$ modes. 
The rotation-induced $B$ mode is proportional to $\a E$, so the rotation angle may be measured from the correlation of $E$ and $B$ modes.
Because temperature is correlated with $E$ modes, the rotation angle may also be measured from temperature-$B$ correlation. 
These effects are the same for any sources of the rotation. 
Using $EB$ and/or temperature-$B$ correlations, 
the uniform polarization rotation angle has been constrained by several groups including 
WMAP \cite{Hinshaw:2013}, BICEP1 \cite{B1rot}, and Planck \cite{P16:rot} (see also Refs.~\cite{Gruppuso:2016,Contaldi:2016}). 
The current best constraints are limited by the accuracy of absolute detector polarization angle calibration.  

Inhomogeneities in pseudoscalar fields and/or PMFs produce anisotropies of the rotation angle \cite{Caldwell:2011,Zhao:2014,De:2013}. 
If the polarization angle is anisotropic, the correlation between $E$ and rotation-induced $B$ modes determined 
at each small patch is also anisotropic. In Fourier space, different Fourier modes of $E$ and $B$ modes correlate. 
Thus the anisotropy of the polarization rotation is extracted through the mode coupling between $E$ and $B$ modes. 
The angular power spectrum of the extracted anisotropic rotation is the four-point correlation of $E$ and $B$ modes, 
and can be reconstructed from the $EBEB$ trispectrum measurement \cite{Kamionkowski:2009}. 
Compared to a uniform rotation, measurements of the anisotropic rotation angle are insensitive
to the accuracy of the overall rotation angle.
There already exist constraints on the cosmic birefringence anisotropies from the CMB. 
Ref.~\cite{Gluscevic:2012qv} presents constraints on 
anisotropies of the cosmic birefringence using the $TBTB$ trispectrum of WMAP7 data,
while Refs.~\cite{Gruppuso:2012,Li:2013,Alighieri:2014yoa,Li:2014,Mei:2014iaa,Pan:2016vai} used two-point correlation. 
The most stringent constraints prior to this paper were published by POLARBEAR \cite{PB15:rot}. 

In this paper, we use a similar method to improve constraints on the 
rotation anisotropies using polarization maps made by BK.

\section{Data and simulations}

We use the same data set described in Refs.~\cite{BKVI} and \cite{BKVIII}: \BK\ maps which coadd all 
data taken up to and including the 2014 observing season---we refer to these as the \bk\ maps.
In this work we use the 150\,GHz $Q/U$ maps. These have a depth of $3.0\ \mu$K-arcmin over an effective area of 
$\sim 395$ deg$^2$, centered on RA 0h, Dec.\ $-57.5$ deg. 

We reuse the standard sets of simulations described in Ref.~\cite{BKVI} and previous papers:
lensed CMB signal-only simulations (denoted by ``lensed-$\Lambda$ Cold-Dark-Matter (lensed-$\Lambda$CDM)'') 
with input lensed maps generated by LensPix \cite{Lewis:2005},
instrumental noise, and dust foreground, each having $499$ realizations.
The details of the CMB signal and noise simulations are given in Sec.~V of Ref.~\cite{B2I}, 
and the dust simulations are described in Sec.~IV A of Ref.~\cite{BKP} and Appendix E of Ref.~\cite{BKVI}. 
In addition, we also generate random fields of anisotropic rotation maps, $\a(\hatn)$, 
on the full sky (where $\hatn$ denotes a position on the sphere) of which the power spectrum is described by 
\al{
	\frac{L(L+1)}{2\pi}C_L^{\a\a} = A_{CB}\times 10^{-4} \qquad \text{[rad$^2$]} \,, \label{Eq:fid:claa}
}
with varying $\ACB$. Since previous constraints on the cosmic birefringence anisotropies are derived based on 
this spectrum, our result can be directly compared with the previous studies 
(see e.g. Refs.~\cite{Caldwell:2011,Gluscevic:2012qv,PB15:rot}). 

The simulated full-sky CMB maps are rotated by $\a(\hatn)$ before beam smoothing according to 
\al{
	[Q'\pm \iu U'](\hatn) = \E^{\pm 2\iu \a(\hatn)}[Q\pm \iu U](\hatn) \,. \label{Eq:rotmap}
}
As described in Ref.~\cite{BKVII}, we simulate observed maps by multiplying the BK14 observing matrix 
with the rotated maps. We denote these maps as ``rotated-$\Lambda$CDM'' simulations. 
The rotated-$\Lambda$CDM, instrumental noise, and dust simulated maps are then combined  
to estimate the transfer function, mean-field bias, disconnected bias, 
and the uncertainties of the power spectrum of reconstructed $\a$.
The reconstructed rotation power can then be compared against lensed-$\Lambda$CDM simulations 
under the null hypothesis to evaluate statistical uncertainties. 

To properly include cosmic variance from $\a$, rotated-$\Lambda$CDM simulations must be used. 
To our knowledge, this has not 
been done in previous papers, in which unrotated simulations are used to calculate uncertainties 
\cite{Gluscevic:2012qv,PB15:rot}). In this paper, we present the test of the null hypothesis 
using the lensed-$\Lambda$CDM simulations to compare our measurements with prior attempts, 
and also show constraints on the anisotropic polarization rotation with the rotated-$\Lambda$CDM simulations. 

\section{Analysis}

The rotation angle anisotropies can be reconstructed from the off-diagonal mode-mode covariance within, 
and between, the $E$ and $B$ modes. An estimator of $\a(\hatn)$ has a quadratic form similar 
to the lensing estimator \cite{Kamionkowski:2009,Yadav:2009}.
The power spectrum of the anisotropic rotation angle $C_L^{\a\a}$
can be obtained by squaring the rotation estimator. 
Here, we describe the method used to reconstruct the anisotropic rotation angle
from the \bk\ polarization maps, to calculate the rotation spectrum, and to
evaluate the amplitudes of the resulting spectra.
The details and verification of our analysis method are described in Ref.~\cite{Namikawa:2016d}. 

Under the flat-sky approximation, the CMB $E$ and $B$ modes are given by 
\al{
	E_{\bl} \pm i B_{\bl} = - \FT{\hatn}{\bl}{f} [Q\pm\iu U](\hatn) \E^{\mp 2\iu\varphi_{\bl}}  \,,
}
where $\varphi_{\bl}$ is the angle of $\bl$ measured from the Stokes $Q$ axis.
From \eq{Eq:rotmap}, the rotated CMB $E$ and $B$ modes are given by \cite{Kamionkowski:2009}
\al{
	\rE_{\bl} &= E_{\bl} + \Int{2}{\bL}{(2\pi)^2} 2\a_{\bL} 
	\notag \\
		&\times [ E_{\bl-\bL}\cos 2(\varphi_{\bl-\bL}-\varphi_{\bl}) 
		+ B_{\bl-\bL}\sin 2(\varphi_{\bl-\bL}-\varphi_{\bl}) ]
	\label{Eq:rotated-E} \\
	\rB_{\bl} &= B_{\bl} + \Int{2}{\bL}{(2\pi)^2} 2\a_{\bL}
	\notag \\
		&\times [ E_{\bl-\bL}\sin 2(\varphi_{\bl-\bL}-\varphi_{\bl})
		- B_{\bl-\bL}\cos 2(\varphi_{\bl-\bL}-\varphi_{\bl}) ]
	\,. \label{Eq:rotated-B}
}

Up to first order in the anisotropic part of $\a$, the rotation-induced off-diagonal elements of 
the covariance are \cite{Kamionkowski:2009}
\al{
	\ave{\rE_{\bl}\rB_{\bL-\bl}}_{\rm CMB} &= w_{\bL,\bl}^\a \a_{\bL} \,, \label{Eq:weight}
}
where $\ave{\cdots}\rom{CMB}$ denotes the ensemble average with a fixed realization of $\a$ and 
the weight function is 
\al{
	w^\a_{\bL,\bl} &= 2\tCEE_\l\cos 2(\varphi_{\bl}-\varphi_{\bL-\bl})
	\,, \label{Eq:weight:a}
}
where $\tCEE_\l$ is the lensed $E$-mode power spectrum. 
The term originating from the lensing $B$ mode is ignored since the improvement of the sensitivity to 
the polarization rotation anisotropies by the inclusion of this term is negligible \cite{PB15:rot}. 
Similar to the lensing reconstruction, the quadratic estimator of $\a$ is constructed as a convolution of 
the $E$ and $B$ modes with the weight function of \eq{Eq:weight:a} \cite{Kamionkowski:2009}. 
The only difference between the reconstruction of $\a$ and the lensing potential, $\phi$, is the weight function. 
Similar to the lensing analysis, we use $E$ and $B$ modes obtained from the matrix-based $E$-$B$ separation technique 
as described in Ref.~\cite{BKVII} to avoid $E$-to-$B$ leakage. 

From the reconstructed $\a$, the rotation spectrum is estimated in the same way as 
the lensing spectrum shown in Ref.~\cite{BKVIII}. The disconnected bias is estimated with 
the realization-dependent method \cite{Namikawa:2016d},
which is more accurate than simulation-based subtraction \cite{Namikawa:2012} 
and also mitigates the off-diagonal elements of covariance \cite{Hanson:2010rp}. 

To quantify the constraints on polarization rotation anisotropies, 
we estimate the amplitude for the reconstructed rotation spectrum \cite{BKVIII}
\al{
	\widehat{A}_{CB} = \frac{\sum_b w_b A_b}{\sum_b w_b} \,, \label{Eq:ACB}
}
where $A_b=C_b/C_b^{\rm fid}$ is an amplitude relative to a fiducial power spectrum at each multipole bin, $b$. 
The coefficients $w_b$ are defined as
\al{
	w_b = \sum_{b'} C^{\rm fid}_b \bR{Cov}^{-1}_{bb'}C^{\rm fid}_{b'} \,, \label{Eq:wb}
}
and the power spectrum covariance $\bR{Cov}_{bb'}$ is estimated from the lensed- and 
rotated-$\Lambda$CDM+noise+dust simulations for evaluating the null hypothesis and constraining $\ACB$, respectively. 
The fiducial rotation spectrum $C_b^{\rm fid}$ corresponds to $\ACB=1$.

In the reconstruction from the rotated-$\Lambda$CDM simulations, 
even after the subtraction of a disconnected bias, 
there exists a non-negligible correction from the secondary contraction at smaller scales \cite{Namikawa:2016d}. 
As detailed in Ref.~\cite{Namikawa:2016d} the secondary contraction of the $EBEB$ trispectrum (N1 term) 
is proportional to the signal, so we include this term for estimating $\hACB$. 
On the other hand, the lensing-induced trispectrum is negligibly small for BK14 data \cite{Namikawa:2016d}.

\section{Reconstructed spectrum}

\begin{figure}[t]
\bc
\includegraphics[width=8.5cm,height=6.5cm,clip]{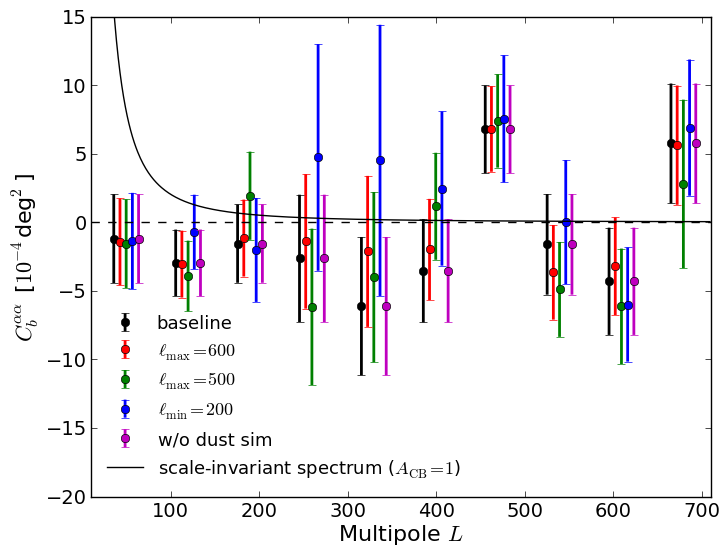} 
\caption{
Angular power spectrum of rotation anisotropies measured from BK14 real data using 
the standard lensed-$\Lambda$CDM+noise+dust simulation to obtain the power spectrum and uncertainties. 
In addition to the baseline analysis we also show cases with different choices of the CMB multipole range 
used for the rotation angle reconstruction and a case without the inclusion of the dust simulation. 
We group the multipoles up to $700$ into $10$ bins.
The solid line shows the scale-invariant spectrum of \eq{Eq:fid:claa} with $\ACB=1$.
}
\label{Fig:aps}
\ec
\end{figure}

Fig.~\ref{Fig:aps} shows the power spectrum of the reconstructed rotation angle from BK14 data.
In the baseline analysis we use CMB multipoles between $\l=30$ and $700$ but remove $B$ modes for multipoles $\l<150$, 
which significantly reduces the large-scale dust foreground contamination (see Ref.~\cite{BKVIII}). 
In addition to the baseline analysis, we also show the cases with different choices of CMB multipole ranges 
used for the rotation angle reconstruction and the case without a dust component. 
We calculate the $\chi^2$ probability-to-exceed (PTE) 
for the baseline analysis and each variant analysis against the null hypothesis. 
For the baseline case the $\chi^2$ PTE is found to be $0.25$. 
The $\chi^2$ PTEs for other cases are in the range between $0.18$ and $0.59$. 
These results indicate that the reconstructed spectrum is consistent with the null hypothesis 
irrespective of the choice of the multipole range and the inclusion of dust in the simulations. 
Fig.~\ref{Fig:aps} indicates that, to constrain the model of \eq{Eq:fid:claa}, 
the largest-scale multipole bin is the most important.
One advantage of BK14 data is the capability of measuring such large scales. 

\begin{figure}[t]
\bc
\includegraphics[width=8.5cm,height=6.5cm,clip]{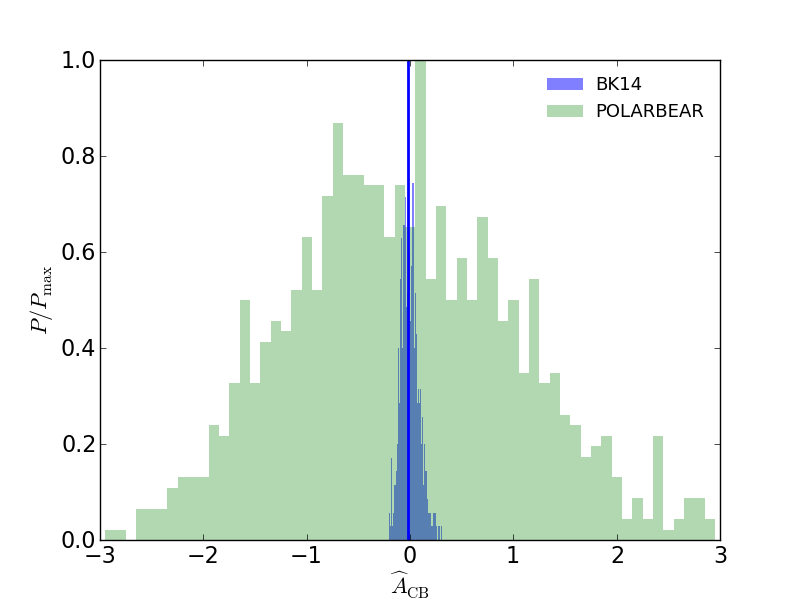} 
\caption{
Histogram of rotation spectrum amplitude $\hACB$ from BK14 data. 
The blue histogram shows the results from the standard $\Lambda$CDM simulations while 
the green histogram shows the POLARBEAR result \cite{FChang}.
The blue vertical line shows the value from the observed spectrum. 
}
\label{Fig:ACB}
\ec
\end{figure}

Fig.~\ref{Fig:ACB} shows the histogram of $\hACB$ for each realization of the null (lensed $\Lambda$CDM+noise+dust) simulations. 
The observed $\hACB$ is shown as the vertical solid line, and is consistent with the null hypothesis.
The rotation spectrum amplitude is estimated from \eq{Eq:ACB}. 
We also show the histogram obtained from the POLARBEAR analysis \cite{FChang} 
which leads to $\ACB<3.1$ at $95\%$ confidence (ignoring the cosmic variance from $\a$). 
The statistical uncertainties for \bk\ are an order of magnitude smaller.
The histogram obtained in this work is skewed because the constraint on $\ACB$ is mostly 
determined by the largest-scale multipoles where the PDF of the power spectrum becomes a chi-squared distribution. 

\section{Cosmological implications}

To obtain a constraint on $\ACB$, we next apply the direct likelihood method of Ref.~\cite{B1} to $\ACB$. 
We run simulations with varying overall amplitude of the 
input scale-invariant spectrum up to $\ACB=1.5$ to
obtain the distribution of $\hACB$ for each value of the input $\ACB$. 
The posterior distribution for the amplitude parameter $\ACB$ is obtained from this direct likelihood 
by assuming a flat prior on $\ACB$ for $\ACB\leq 1.5$.
The resulting constraint is $\ACB\leq 0.33$ at $95\%$ confidence
and is the best constraint on cosmic birefringence anisotropies to date.

Using \eq{Eq:scaleinv}, this $\ACB$ constraint can be translated into constraints on coupling between 
axionlike particles and photons
\al{
	g_{a\gamma} \leq \frac{7.2\times 10^{-2}}{H_I} \,. 
}
This is at least an order of magnitude better than the constraint from 
Ref.~\cite{Pospelov:2009} which obtains $g_{a\gamma}H_I\alt 1$.

The constraint above leads to implications for axionlike particles with a small mass as discussed in e.g., Ref.~\cite{Marsh:2016}. 
In general, if axionlike particles have a mass, $m_a$, the field value perturbation starts to oscillate 
when the Hubble friction becomes inefficient as similar to the uniform unperturbed value. 
The change of the field value in \eq{Eq:alpha}, and equivalently the polarization rotation angle, 
is significantly suppressed after the oscillation. 
Thus, the polarization rotation anisotropies are generated if the oscillation of the axionlike particles 
starts after recombination ($t=t_{\rm rec}$). 
Since the time of the transition to oscillation is given by $H(t_{\rm osc})\sim m_a$, 
the mass range of the axionlike particles is $m_a=10^{-33}-10^{-28}$ eV where the lower limit comes 
from $m_a\sim H_0$ and the upper bound comes from $m_a\sim H(t_{\rm rec})$. 
The string axion generally predicts such a mass spectrum.
According to Fig.~2 of Ref.~\cite{Ringwald:2014}, 
the constraint on $g_{a\gamma}$ presented above is much tighter than other experiments at $m_a=10^{-33}-10^{-28}$
if the tensor-to-scalar ratio is $r\sim 0.01$. 

The coupling constant is related to the decay constant $f_a$ as $g_{a\gamma}=(\alpha/2\pi) C_{a\gamma}/f_a$ where $\alpha$ 
is the fine structure constant and $C_{a\gamma}$ is a model-dependent dimensionless coupling. 
The typical value of $C_{a\gamma}$ is $\mC{O}(1)$. 
The value of the decay constant in string theory models is typically $f_a\sim 10^{16}$ GeV 
(e.g., the model-independent axion in heterotic string theory and M-theory axiverse) but could be $f_a\alt M_{\rm pl}$ 
(type IIB theory) with $M_{\rm pl}$ denoting the Planck energy scale \cite{Marsh:2016}. 
Our constraint tightens the allowed region of $f_a$ for string axions with a mass within the above mass range. 
For example, if $r\sim 0.01$ and $C_{a\gamma}\sim 1$, 
we obtain $H_I\sim 10^{-5}M_{\rm pl}$ and the allowed range becomes $10^{-6}\alt f_a/M_{\rm pl}\alt 1$. 
In the near future, measurement of polarization rotation from 
CMB-S4\footnote{\url{https://cmb-s4.org/CMB-S4workshops/index.php/Main_Page}} 
would further improve the lower bound 
by $\sim 4-5$ orders of magnitude compared to our results, and significantly constrain $f_a$. 

Following Refs.~\cite{De:2013,Pogosian:2014,PB15:rot}, we can also convert the above upper bound to 
the amplitude of the PMFs. The above result constrains 
the strength of the scale-invariant PMFs smoothed over 1Mpc to $B_{1Mpc}\leq 30$nG, 
which is roughly three times better than that obtained from the previous best constraints on 
the polarization rotation 
(note that other statistics such as the PolarBear $BB$ spectrum at high $\ell$ can further tighten the magnetic-field constraint 
compared to the trispectrum constraint presented here). 

Note that a $BB$ spectrum is also generated by the anisotropies of the cosmic birefringence through
conversion from $E$ to $B$ modes. 
The BK14 $BB$ spectrum is, however, less sensitive to cosmic polarization rotation anisotropies than $C_L^{\a\a}$, 
and the upper bound on the cosmic polarization rotation anisotropies using the $BB$ spectrum is much larger than $A_{CB}\leq 0.33$.
In other words, the results in this paper also rule out significant contributions from cosmic 
birefringence to BK14's main $BB$ results, a possibility raised by Ref.~\cite{Li:2015vea}.

\section{Discussion}

The BK14 data have been extensively searched for possible systematics in previous publications 
in the power spectrum and lensing trispectrum.
To further test potential systematic contamination in the measured rotation spectrum,
we perform rotation reconstruction on differenced (``jackknife'') maps and check whether they are 
consistent with null (see Ref.~\cite{B2III} for the details of the jackknife maps). 
Table \ref{table:PTE} shows the $\chi^2$ PTE for 
these jackknife tests. The jackknife spectra show no evidence of spurious signals.

\begin{table}
\bc
\caption{
Probability to exceed a $\chi^2$ statistic for the jackknife tests (see Ref.~\cite{B2III} for 
definitions of these jackknife splits). 
}
\label{table:PTE}
\begin{tabular}{l|cc|ccc} \hline
Deck                    & 0.822 \\
Scan Dir                & 0.856 \\
Tag Split               & 0.064 \\
Tile                    & 0.285 \\
Phase                   & 0.776 \\
Mux Col                 & 0.383 \\
Alt Deck                & 0.567 \\
Mux Row                 & 0.715 \\
Tile/Deck               & 0.964 \\
Focal Plane inner/outer & 0.375 \\
Tile top/bottom         & 0.924 \\
Tile inner/outer        & 0.248 \\
Moon                    & 0.375 \\
A/B offset best/worst   & 0.194 \\
\hline
\end{tabular}
\ec 
\end{table}

Galactic dust contamination affects the rotation spectrum measurement by producing 
an additional disconnected bias and trispectrum induced by dust non-Gaussianity. 
While a thorough estimation of these two effects requires a reliable non-Gaussian dust simulation, 
the following evidence demonstrates that our rotation spectrum measurement is not significantly 
affected by Galactic dust:
\bi
\item We estimate the rotation spectrum by repeating the simulations with no dust and 
show that the change of the spectrum is negligible compared to the statistical uncertainties. 
This means that the additional disconnected bias by the Gaussian dust component is negligible.
Since the power of the non-Gaussian dust is comparable to that of the Gaussian dust, 
the impact of the non-Gaussian dust on the disconnected bias would also be negligible.
\item To test the possible impact of dust we tighten the cut on large-scale $B$ modes
from $150<\ell$ to $200<\ell$.
The results remain consistent with the null hypothesis. 
\item The dust could also lead to nonzero cross-power between the lensing and rotation maps. 
We cross-correlate the reconstructed rotation angle 
with the reconstructed lensing maps from BK14 shown in our lensing paper \cite{BKVIII} 
and also with the public Planck 2015 lensing maps \cite{P15:phi}. 
The $\chi^2$ PTEs of these cross-spectra are $0.75$ 
for $\alpha\times\kappa^{\rm BK14}$ and $0.63$ for $\alpha\times\kappa^{\rm P15}$.
We find the cross-spectrum to be consistent with zero. 
\ei
These negative results suggest that the dust foreground contamination is not significant 
in the reconstructed rotation spectrum. 

In our analysis the overall polarization angle is calibrated by minimizing 
the $TB$ and $EB$ spectra \cite{B1rot,Keating:2013,BKVI,PB15:rot}. 
However, limited accuracy of relative detector polarization calibration can also affect rotation spectrum measurements. 
To test this we generate a set of signal-only time-ordered-data (TOD) simulations 
in which the baseline detector polarization angles are offset
according to measured values for Keck 2014 data (see Ref.~\cite{BKIV} for details). 
We then coadd them to maps using the nominal detector polarization angles. 
We repeat the analysis replacing the standard $\Lambda$CDM signal with this simulation, 
finding that the change in the reconstructed power spectrum is $<1\%$ of 
the $1\sigma$ statistical uncertainty in all band powers. 
Even if we repeat the analysis using the simulation where the offsets from nominal are multiplied by $5$, 
the change in the reconstructed spectrum is still $\sim 1\%$ of the $1\sigma$ statistical error. 
We therefore conclude that the systematic errors due to relative detector polarization angle offsets are negligible in our analysis.

\section{Conclusion}

We present measurement of anisotropies of the CMB polarization rotation angle using BK14 data 
and find that the spectrum is in agreement with the null hypothesis 
(the standard $\Lambda$CDM prediction). 
The $95\%$ upper bound on the amplitude of the scale-invariant 
rotation spectrum relevant to the inflationary scenario is 
$0.33\times 10^{-4} [{\rm rad^2}]=0.11\,{\rm deg}^2$ which is approximately 
ten times better than the best previous result~\cite{FChang}. 
The measured rotation spectrum is used to constrain 
cosmic birefringence from axionlike particles and Faraday rotation of PMFs. 
The constraint presented in this paper tightens the allowed range of the coupling constant 
for axionlike particles with $m_a=10^{-33}-10^{-28}$. 
At this mass range, the CMB polarization rotation measurement is the best avenue to probe 
the axionlike particles, and in the near future CMB-S4 will further tighten the allowed parameter space. 
We test systematics in the measured rotation spectrum by 
1) performing jackknife null tests,
2) cross-correlating with gravitational lensing maps, and 
3) evaluating the effect of relative rotation angle offsets between detectors, finding no spurious signals. 

The anisotropic rotation angle is a unique probe of parity-violating models, 
and its measurement is important to test new physical theories of the early Universe. 
Future CMB experiments such as the BICEP Array, Advanced ACT, CMB-S4, LiteBIRD, Simons Array, and SPT-3G
will measure rotation angle anisotropies more precisely. 

\begin{acknowledgments}
The \keckarray\ project has been made possible through support from the National Science Foundation 
under Grants ANT-1145172 (Harvard), ANT-1145143 (Minnesota) \& ANT-1145248 (Stanford), and from the 
Keck Foundation (Caltech). The development of antenna-coupled detector technology was supported
by the JPL Research and Technology Development Fund and Grants No.\ 06-ARPA206-0040 and 10-SAT10-0017 
from the NASA APRA and SAT programs. The development and testing of focal planes were supported
by the Gordon and Betty Moore Foundation at Caltech. Readout electronics were supported by a Canada 
Foundation for Innovation grant to UBC. The computations in this paper were run on the Odyssey cluster
supported by the FAS Science Division Research Computing Group at Harvard University. The analysis 
effort at Stanford and SLAC is partially supported by the U.S. Department of Energy Office of Science.
We thank the staff of the U.S. Antarctic Program and in particular the South Pole Station without 
whose help this research would not have been possible. Most special thanks go to our heroic 
winter-overs Robert Schwarz and Steffen Richter. We thank all those who have contributed past efforts 
to the \bicep--\keckarray\ series of experiments, including the \bicepone\ team.
We thank Chang Feng for providing the histogram data shown in Ref.~\cite{PB15:rot}. 
\end{acknowledgments}

\bibliographystyle{apsrev}
\bibliography{cite_exp,cite_general}

\end{document}